\documentclass[summary]{emts2019}

\usepackage{amsmath,amssymb,epsfig,amsfonts}
\usepackage{latexsym,graphicx,graphics}
\usepackage{bm}

\newcommand{\abs}[1]{\left| #1 \right|}

\newcommand{\bb}[1]{ \bm #1}

\newcommand{\rv}{\bb r}

\newcommand{\rhov}{\bb{\rho}}
\newcommand{\rhovp}{\bb{\rho'}}

\newcommand{\Eq}[1]{Eq.~(\ref{#1})}

\newcommand{\Gst}{G^{\mbox{st}}}

\newcommand{\unit}[1]{\bb{\hat{#1}}}

\title{Breach of symmetries in rotating arrays and metamaterials observed in their rest frame}

\author{Ido Kazma and Ben Z. Steinberg*\affref{ref1}
  }

\affiliation{%
  \aff{ref1}{School of EE, Tel-Aviv University, Tel-Aviv 69978, Israel}
}

\begin{document}

\maketitle

\begin{abstract}
 Polarizability theory and discrete dipole approximation in a rotating medium rest-frame is developed and discussed. The analysis is based on a rigorous rotating medium Green's function, and is used to study the effect of rotation on various arrays and metamaterials. The non-reciprocal electrodynamics induced by the rotation is exposed and studied. Although it can be interpreted in terms of the multiplicity of Sagnac interference loops inside the structure, the associated rotation footprint exhibits new effects of the structure parameters not previously reported in conventional Sagnac effect. The resulting non-reciprocity will also be demonstrated and discussed.

\end{abstract}

\section{Introduction}

We study the effect of rotation on complex arrays and metamaterials and/or periodic structures consisting of electrically small scatterers, as observed in the rotating medium's rest-frame. To observe the rotation footprint in a rigorous and systematic framework, we use 2D Green's function theory in slowly rotating medium developed in \cite{BZSG}. We examine the effect of rotation on the polarizability of electrically small objects, and developed closed form expressions for the case of dielectric cylinders. These tools pave the way for a systematic study of the electrodynamics of rotating structures and metamaterials as observed in their rest frame of reference.

The ED of rotating structures as observed in the laboratory (inertial) frame of reference has been investigated extensively. Pioneering studies can be found, e.g. in \cite{VanBladel1,DeZutter1,VanBladelBook}. The main difficulty in their generalization and application stems from the \emph{moving boundaries}. To compare, when a rigid structure that rotates at an angular (radian) velocity $\bb{\Omega}$ is observed in its rest (and non-inertial) frame of reference $\mathcal{R}^{(\Omega)}$, its boundaries are \emph{stationary}.
We set now the following terminology. Here and henceforth, we always observe the electromagnetic system in its rest frame of reference, i.e. in a frame where its boundaries do not move. Thus, by ``a problem in $\mathcal{R}^{(0)}$'' we mean a system that does \emph{not} rotate, observed in the inertial (laboratory) frame of reference. Likewise, by ``a problem in $\mathcal{R}^{(\Omega)}$'' we mean a system that rotates rigidly at an angular velocity $\bb{\Omega}=\unit{z}\Omega$, observed in the non-inertial frame of reference where it appears at rest.

It has been shown that if the problem in $\mathcal{R}^{(0)}$  is described by the scalar permittivity and permeability $\epsilon(\rhov)=\epsilon_0\epsilon_r(\rhov),\,\mu(\rhov)=\mu_0\mu_r(\rhov)$, then in the limit of slow rotation the problem in $\mathcal{R}^{(\Omega)}$ is still governed by the conventional set of Maxwell's equations (ME), where the rotation is manifested only via the modified constitutive relations \cite{Shiozawa}
\begin{subequations}
\begin{align}
\bb{D}&=\epsilon\bb{E}-c^{-2}(\bb{\Omega}\times\rv)\times\bb{H}\label{eq1a}\\
\bb{B}&=\mu\bb{H}+c^{-2}(\bb{\Omega}\times\rv)\times\bb{E}\label{eq1b}
\end{align}
\end{subequations}
where $c$ is the speed of light in vacuum. For homogeneous $\epsilon_r,\mu_r$ and for $z$-independent excitations, the resulting set of ME can be separated to independent TE and TM fields, and rigorous 2D Green's function can be developed \cite{BZSG}. The reader is referred to \cite{BZSG} for details. Below we use a compact version of this Green's function to discuss interference and diffraction patterns for periodic structures.

\section{Rotating medium Green's function}

For homogeneous $\epsilon,\mu$, the 2D problem can be rigorously separated into decoupled TE and TM polarizations, in both the complete electromagnetic field can be derived from the corresponding $z$-directed field satisfying a modified Helmholtz equation \cite{BZSG}
\begin{equation}\label{eq2}
[\nabla_t^2+k_0^2n^2]F_z-2ik_0^2\frac{\Omega}{\omega}\partial_\theta\,F_z=S
\end{equation}
where $k_0=\omega/c$, $F_z=H_z\, (E_z)$ for TE (TM), and $n^2=\epsilon_r\mu_r$. Here $S=S^{\mbox{\tiny TE}}=-i\omega\epsilon J_z^M-i\frac{\omega\Omega}{c^2}\bb{\rho}\cdot\bb{J}_t-\unit{z}\cdot\nabla_t\times\bb{J}_t$, or $S=S^{\mbox{\tiny TM}}=-i\omega\mu J_z+i\frac{\omega\Omega}{c^2}\bb{\rho}\cdot\bb{J}_t^M+\unit{z}\cdot\nabla_t\times\bb{J}_t^M$. Hence, for any $S$, the field can be obtained via the scalar Green's function, defined as the response to the current $\bb{J}=\unit{z}I\delta(\rhov-\rhovp)$,
\begin{equation}\label{eq3}
[\nabla_t^2+k_0^2n^2]G-2ik_0^2\frac{\Omega}{\omega}\partial_\theta\,G=
-\frac{1}{\rho'}\delta(\rho-\rho')\delta(\theta-\theta').
\end{equation}
This Green's function is given by
\begin{equation}\label{eq4}
G(\rhov,\rhovp)=\frac{i}{4}\!\!\!\sum_{m=-\infty}^\infty \!\!\!\! J_m(k_0 n \gamma_m\rho_<)H_m^{(1)}(k_0 n \gamma_m\rho_>)e^{im(\theta-\theta')}
\end{equation}
where $\gamma_m=\sqrt{1+2m\Omega/(\omega n^2)}$ and $\rho_\gtrless=\substack{\mbox{\tiny max} \\ \mbox{\tiny min}}(\rho,\rho')$. To overcome the difficulties associated with the slowly converging series, an approximation has been suggested \cite{BZSG}
\begin{equation}\label{eq5}
G(\rhov,\rhovp)\approx G_{\mbox{\tiny app}}=\Gst(\rhov,\rhovp)e^{ik_0(\Omega/c)\unit{z}\cdot(\rhovp\times\rhov)}
\end{equation}
where $\Gst(\rhov,\rhovp)=\frac{i}{4}H_0^{(1)}(k_0n|\rhov-\rhovp|)$ is the 2D Green's function of a homogeneous medium in $\mathcal{R}^0$.
It is worth noting that the approximation in \eqref{eq5} possesses the following properties.
First, note that the rotation is manifested only in the term $e^{ik_0(\Omega/c)\unit{z}\cdot(\rhovp\times\rhov)}$, that \emph{does not} depend on the medium properties.
Then, it is easily shown that the Sagnac phase shift as predicted by this Green's function, is independent of the medium refraction index; the phase shift between a co-rotating and counter-rotating light beams enclosing an area $S$, as observed in $\mathcal{R}^{(\Omega)}$, is $\Delta\phi=4k_0\Omega S/c$. This is consistent with previous studies of Sagnac effect \cite{Post}.
Furthermore,
according to the basic theory of PDEs, the Green's function \emph{singularity} at $\rhov\rightarrow\rhovp$ is determined only by the highest order derivative term in the equation. Thus, the singularity of $G$ and $\Gst$ should be the same [see the governing Helmholtz operator in \eqref{eq2}]. This indeed is preserved by the approximated Green's function above.

The properties above imply that \eqref{eq5} is in fact a \emph{uniform} approximation of our Green's function, that encapsulates all the essential physics contained in the exact expression \eqref{eq4}. The advantages of the representation in \eqref{eq5} are clear; the need to sum slowly converging series (especially in the near field) has been alleviated. Furthermore, the simplicity of this expression enables one to investigate interference and diffraction phenomena by using essentially the same tools used for interference in $\mathcal{R}^{(0)}$.

\section{Polarizability theory for rotating medium}

In the framework of polarizability theory and discrete dipole approximation (DDA), the response of a set of $N$ electrically small scatterers to an incident field $\bb{E}^{\mbox{\tiny inc}}(\rv)$ is governed by
\begin{equation}\label{eq6}
\bb{I}_n-i\omega\mu_0 \bb{\alpha}_n\sum_{m\ne n}G(\rv_n,\rv_m) \bb{I}_m=
\bb{\alpha}_n\bb{E}^{\mbox{\tiny inc}}(\rv_n),\quad n=1,\ldots N
\end{equation}
where $\bb{I}_n=-i\omega \bb{p}_n$ is the polarization current induced in the $n$-th scatterer and $\bb{p}_n$ is its dipole-moment response, $\bb{\alpha}_n$ is its polarizability, and $G(\rv_n,\rv_m)$ is the medium's Dyadic Green's function. In the above, $\bb{\alpha}$ is defined via $\bb{I}=\bb{\alpha} \bb{E}^{L}$ where $\bb{E}^L$ is the local field; the electric field in the scatterer location, in the absence of the scatterer. The appropriate Green's function can be obtained using the formulation discussed in the previous section. However, the polarizabilities in $\mathcal{R}^{(\Omega)}$ still need to be studied. Towards this end, we note that by using the volumetric Method of Moments (MoM) for the TM polarization, one in fact needs to solve the following integral equation for the field in a single dielectric scatterer,
\begin{equation}\label{eq7}
E_z(\rhov)=E_z^{\mbox{\tiny inc}}(\rhov)+k_0^2
\int_{S'}\!\!dS'\, C(\rhovp)\,E_z(\rhovp)\,G(\rhov,\rhovp)
\end{equation}
where $S$ is the scatterer cross-section area, and $C(\rhov)=n^2-n_b^2$ is the contrast function of the dielectric scatterer as measured in $\mathcal{R}^{(0)}$, and $G(\rhov,\rhovp)$ is our rotating medium Green's function in \Eq{eq4} or \Eq{eq5}. Since the scatterer is electrically small, one can approximate its internal electric field by a constant $E_z$, resulting with the following equation,
\begin{equation}\label{eq8}
E_z=E_z^{\mbox{\tiny inc}}(\rhovp)\left(1-k_0^2 C s^{-1}\mathcal{I}_2\right)^{-1}
\end{equation}
where $\rhovp$ is the scatterer's center location, $s$ is the scatterer cross-section area, $C=n^2-n_b^2$ is the (constant) contrast, and $\mathcal{I}_2$ is the following double integral
\begin{equation}\label{eq9}
\mathcal{I}_2=\int_B dS\int_B dS' G(\rhov,\rhovp)
\end{equation}
where $B$ is a box function covering $s$ whose typical dimension is much smaller then the wavelength. Hence, one may use the uniform Green's function in \Eq{eq5} and approximate the exponent by its Taylor series. The result is
\begin{align}\label{eq10}
\mathcal{I}_2(\Omega)&=\int_S ds\int_S ds' \, \Gst(\rhov,\rhovp)\left[1+\right.
i\frac{\Omega\omega}{c^2}\unit{z}\cdot(\rhovp\times\rhov)\nonumber\\
\\
                     &-
\frac{1}{2}\left(\frac{\Omega\omega}{c^2}\right)^2\abs{\unit{z}\cdot(\rhovp\times\rhov)}^2
                     \left. +\cdots\right].\nonumber
\end{align}
Clearly, $\Gst$ is even with respect to interchanging the roles of $\rhov$ and $\rhovp$, while the first order term in $\Omega$ inside the square brackets is odd. Since both $\rhov,\rhovp$ span the same area under the integration, it terns out that the first order term in $\Omega$ vanishes. Therefore, the effect of rotation on $\mathcal{I}_2$ is only second order in $\Omega$ and under the slow rotation assumption it can be neglected. Furthermore, since the polarization current inside the scatterer is nothing but $-i\omega\epsilon_0\chi E_z$, where in our case $E_z$ is given in \Eq{eq8}, it turns out that the effect of rotation on the particle polarizability is also second order in $\Omega$ and can be neglected in practical rotation rates. While the analysis above is carried only for the TM polarization, the same consequences hold also for the TE case (although the pertaining derivation is more complicated).

\subsection{Array Construction}\label{ArrayConstruction}

 The number of different Sagnac loops $N_{\mbox{\tiny SL}}$ inside an array of electrically small scatterers may serve as a qualitative estimate for its potential sensitivity to rotation, as shown qualitatively in Fig.~ref{figure5}. $N_{\mbox{\tiny SL}}$ equals the number of \emph{ordered sets} that can be chosen out of $N$ elements, excluding the empty set, the set of a single element, and the set of two elements. Hence we have
\begin{figure}[ht]
\centering\includegraphics[scale=0.35]{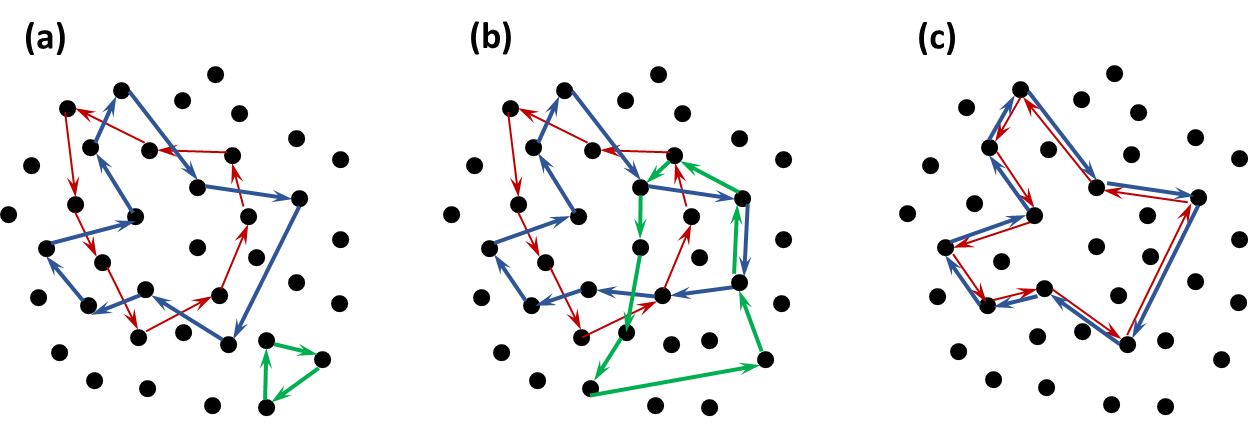}
\caption{Sagnac loops defined by a set of $N$ point-scatterers. (a) Different Sagnac loops. (b) Different Sagnac loops that share at least one common point-scatterer. (c) Different Sagnac loops that share all their point scatterers.}
\label{figure5}
\end{figure}
\begin{align}\label{eq14}
N_{\mbox{\tiny SL}}&=N(N-1)(N-2)+N(N-1)(N-2)(N-3)+\nonumber\\
    &+\cdots+N!
=\sum_{n=3}^N\frac{N!}{(N-n)!}.
\end{align}
For $N\gg 1$
$N_{\mbox{\tiny SL}}\approx eN!-(N^2+1)$ (recall the series expansion of $e^1$).
We now slightly refine our physical picture. Assume that the structure is excited by some source, and the response of the $n$-th point scatterer is $E_n\equiv E_z(\rhov_n)$. The sensitivity of $E_n$ to rotation is due to the total of different Sagnac loops that share the $n$-th point scatterer. This is due to the fact that different loops experience different $\Omega$-dependent phase-shifts, and they interfere at their common point-scatterers. Examples are shown in Fig.~\ref{figure5}(b)-(c). Thus, we look for $N_{\mbox{\tiny SL}1}$: the number of different SL that share at least one point-scatterer. This is nothing but
$N_{\mbox{\tiny SL}1}=N(N-1)_{\mbox{\tiny SL}}$.
Clearly, this number increases very fast with $N$. It represents, in fact, the number of waves with random phases that interfere at each point scatterer. Therefore, we anticipate that it may serve as a measure of the potential sensitivity to rotation.

\subsubsection{Random vs. periodic arrays}\label{RandPer}

\begin{figure}[ht]
\centering\includegraphics[scale=0.45]{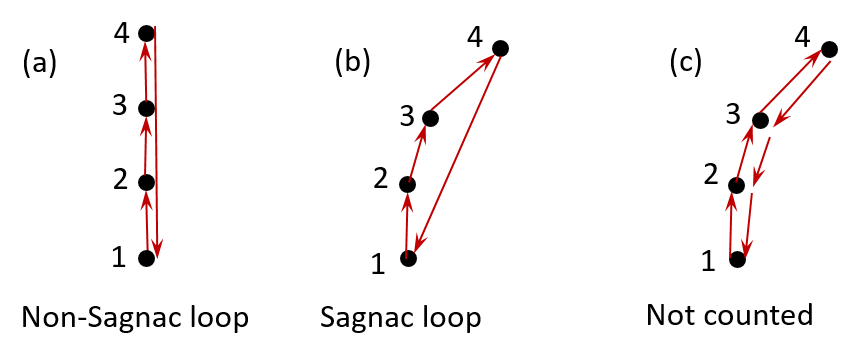}
\caption{A closed loop that consists of four point scatterers, with the ordered interactions $1\Rightarrow2\Rightarrow3\Rightarrow4\Rightarrow1$. (a) All four scatterers reside on a straight line, hence the enclosed area is zero. (b) The four scatterers cannot be aligned along a single straight line, hence the enclosed area does not vanish. (c) Scattering events ordered as $1\Rightarrow2\Rightarrow3\Rightarrow4\Rightarrow3\Rightarrow2\Rightarrow1$, are never counted in our analysis since they have repeated elements.}
\label{figure6}
\end{figure}

We would like to maximize the number of Sagnac loops created by the array of $N$ scatterers.
Consider the ordered subset of $M$ non-repeated scatterers, $2<M\le N$, with the sequence of scattering events $1\Rightarrow 2\Rightarrow\ldots \Rightarrow M$ such as those shown in Fig.~\ref{figure5}. A necessary condition for the existence of Sagnac effect in this series of events is that the enclosed area does not vanish. This is satisfied only if the $M$ scatterers \emph{do not reside on a single straight line}, as schematized in Fig.~\ref{figure6}. It is clear that the condition above is satisfied for every $2<M\le N$ iff it is satisfied for $M=3$. Thus, for a given array, it is sufficient to verify that \emph{every} set of 3 scatterers does not reside on a straight line (or equivalently: define a triangle of non-vanishing area.)

A periodic array cannot satisfy the condition discussed above. A natural choice is a random array since in this case the probability to find three points on the same line is zero. Other structures that may fulfill the condition are quasi-periodic arrays of various types, or arrays that are not periodic nor quasi-periodic, but still not random, such as the Golden Angle (GA) spiral \cite{Cao_Vogel1}.

\section{Examples}

We have used the discrete dipole approximation to compute interference patterns in rotating arrays. Three types of arrays are considered. A rectangular periodic array, a GA spiral, and a square random array. In all the examples below the background material is vacuum. The dielectric cylinders are are made of Si ($\epsilon_r=11.4$), and their radii is $\lambda/50$ where $\lambda$ is the vacuum wavelength. The number of cylinders is 2500 ($50\times 50$ in the rectangular array), and the arrays are excited by a point source located at the array center.
We choose, for convenience, $\lambda=1\mu m$. The inter-cylinder distance is set to $2\lambda$ in the rectangular array. In the GA spiral it is not possible to set a fixed single value that holds for all inter-particle distances. Here, we tuned the GA scaling parameters such that the minimal inter-particle distance is $2\lambda$.
In the square random array we used a uniform distribution between $[-75,75]$ along both $(x,y)$ to obtain the cylinder's locations. To avoid clustering we set the minimal distance between cylinders to $d_{\mbox{\tiny min}}=2\lambda$.

\begin{figure}[htpb]
\hspace*{-0.35in}
\centering\includegraphics[scale=0.36]{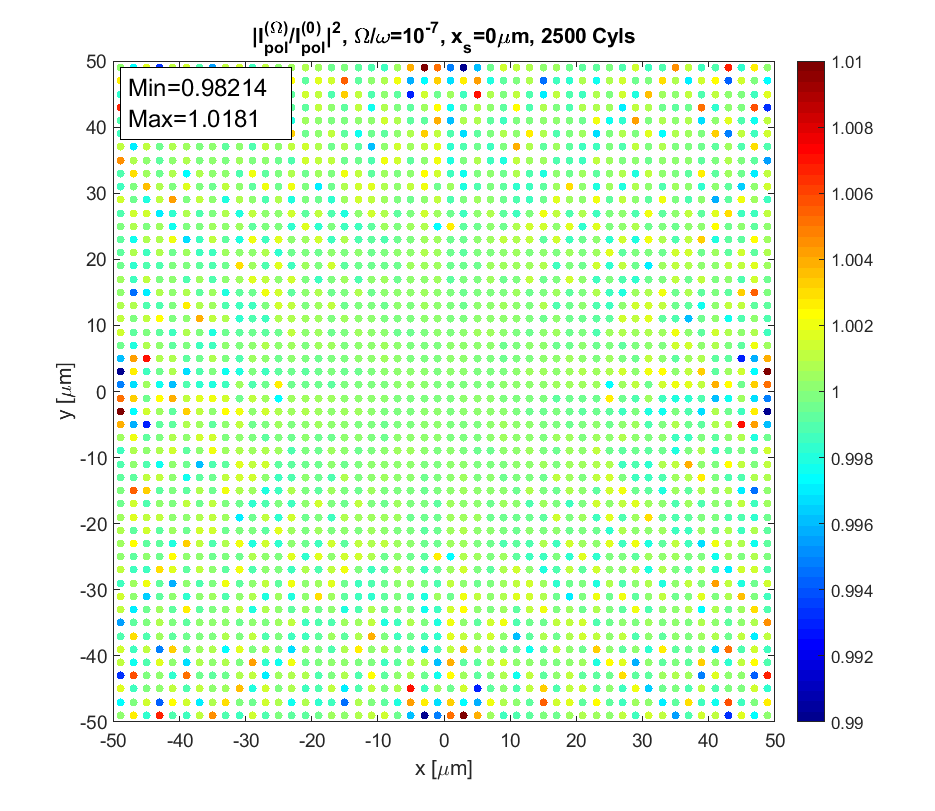}
\caption{Excitation in a rotating rectangular array.}
\label{figSQ1}
\end{figure}

Figure \ref{figSQ1} shows the polarization current excited in the cylinders in a rotating rectangular array, normalized to the corresponding polarization currents in a stationary array, i.e.
 $\abs{I^{(\Omega)}_{\mbox{\tiny pol}}/I^{(0)}_{\mbox{\tiny pol}}}$, for array rotating around its center at $\Omega/\omega=10^{-7}$. The minimal and maximal values of this ratio across the entire array were $[0.98214,\, 1.0181]$, spanning a range of about $3.6\%$. These extremal values are printed in the square box in the top left corner of the figure. To get a better grasp of the pattern, we increased contrast by scaling the colormap only between $(0.99,1.01)$, so some cylinders color is saturated. For the increased rotation rates $\Omega/\omega=10^{-6}$ and $\Omega/\omega=5\times10^{-6}$, the minimum and maximum rations changed to $[0.8336,\, 1.1943]$ and $[0.3953,\, 2.2943]$ respectively. We have repeated these simulations for a rotation axis shifted leftward by $1000\mu m$, and observed that the pattern is independent of the rotation axis location. It is well known that the Sagnac effect is independent of the rotation axis. Therefor this observation supports our understanding that the physical processes that dominate the pattern dependence on $\Omega$ are series of scattering events that create closed Sagnac loops. This observation is also pleasing from the practical point of view, as one would like to design rotation sensors whose output does not depend on the axis location. We have repeated the simulation also for a rotation in the opposite direction; this inversion is manifested by an inversion map of the pattern: $(x,y)\mapsto(-x,-y)$.

\begin{figure}[htpb]
\hspace*{-0.35in}
\centering\includegraphics[scale=0.36]{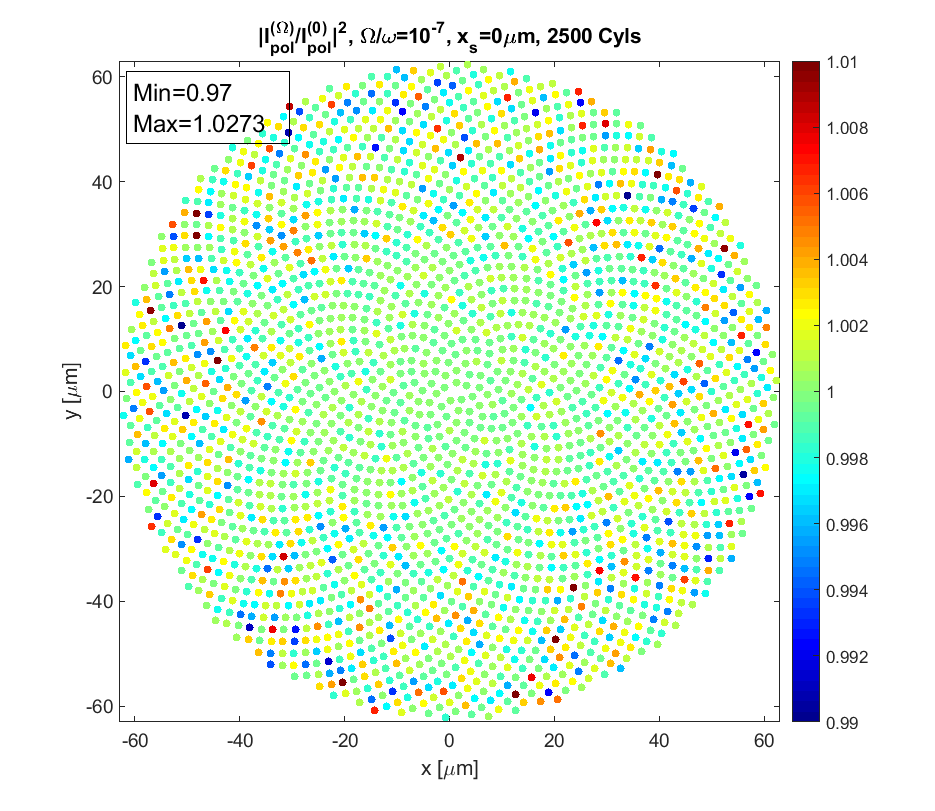}
\caption{Excitation in a rotating GA array.}
\label{figGA1}
\end{figure}

\begin{figure}[htpb]
\hspace*{-0.35in}
\centering\includegraphics[scale=1.3]{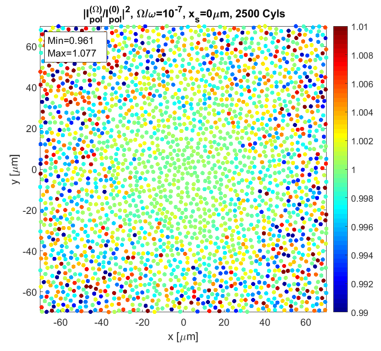}
\caption{Excitation in a rotating GA array.}
\label{figRAND1}
\end{figure}

Figure~\ref{figGA1} shows the same as Fig.~\ref{figSQ1}, but for the GA spiral array with 2500 cylinders. For a convenient comparison to the square array case, the colormap scaling in each pallets here is identical to the colormap of the corresponding pallet in Fig.~\ref{figSQ1}. the maximal and minimal values observed in each case are again shown in the top left corner of pallets. It is seen that they span a significantly larger range. Hence, sensitivity to rotation has been increased. This observation is consistent with our qualitative discussion in Sec.~\ref{RandPer}. We note that also here the picture is independent of the rotation axis location. However, here $\Omega\mapsto -\Omega$ is \emph{not} manifested by the inversion $(x,y)\mapsto(-x,-y)$, since the GA array possesses no geometrical symmetry and no periodicity and the wave dynamics is non-reciprocal due to rotation.
Figure~\ref{figRAND1} shows the same but for the random array. Again, sensitivity to rotation has been increased, reassuring the physical arguments presented in Sec.~\ref{RandPer}. As with the previous example, the picture is independent of the rotation axis, and does not exhibit any symmetry under the change $\Omega\mapsto -\Omega$.

\section{Acknowledgements}

The author BZS gratefully acknowledges fruitful discussions with Prof.~Hui Cao at Yale.

\end{document}